\documentclass[journal]{IEEEtran}
\usepackage{cite}

\ifCLASSINFOpdf
  \usepackage[pdftex]{graphicx}
\else
  \usepackage[dvips]{graphicx}
\fi
\usepackage[cmex10]{amsmath}
\usepackage{amssymb}
\usepackage{algorithmic}

\usepackage{array}

\ifCLASSOPTIONcompsoc
 \usepackage[caption=false,font=normalsize,labelfont=sf,textfont=sf]{subfig}
\else
 \usepackage[caption=false,font=footnotesize]{subfig}
\fi
\usepackage{mathtools}
\usepackage{framed}
\usepackage{wasysym}
\usepackage{hyperref}

\usepackage{tikz}
\usepackage{textcomp}

\newtheorem{theorem}{Theorem}
\newtheorem{lemma}{Lemma}
\newtheorem{corollary}{Corollary}

\newcommand{\pmin}{p'}
\newcommand{\pmax}{p''}
\newcommand{\Pmin}{P'}
\newcommand{\Pmax}{P''}
\newcommand{\Pcor}{\mathcal{P}^{=}}
\newcommand{\Pmu}{\mathcal{M}_{\mu}}
\newcommand{\dint}{\mathrm{d}}
\newcommand{\D}[3]{D_{#1}(#2 \Vert #3)}

\newcommand{\RN}[1]{%
  \textup{\uppercase\expandafter{\romannumeral#1}}%
}

\let\originalleft\left
\let\originalright\right
\renewcommand{\left}{\mathopen{}\mathclose\bgroup\originalleft}
\renewcommand{\right}{\aftergroup\egroup\originalright}

\newcommand\copyrighttext{%
  \footnotesize \textcopyright 2016 IEEE. Personal use of this material is permitted.
  Permission from IEEE must be obtained for all other uses, in any current or future
  media, including reprinting/republishing this material for advertising or promotional
  purposes, creating new collective works, for resale or redistribution to servers or
  lists, or reuse of any copyrighted component of this work in other works.
  DOI: \href{https://doi.org/10.1109/TSP.2016.2600505}{10.1109/TSP.2016.2600505}}
\newcommand\copyrightnotice{%
\begin{tikzpicture}[remember picture,overlay]
\node[anchor=south, yshift=3pt] at (current page.south) {\fbox{\parbox{\dimexpr\textwidth-\fboxsep-\fboxrule\relax}{\copyrighttext}}};
\end{tikzpicture}%
}

\begin{document}
%
\title{Old Bands, New Tracks---Revisiting the Band Model for Robust Hypothesis Testing}
%
%
%

\author{Michael~Fau\ss{}~\IEEEmembership{Student Member,~IEEE,}
        and~Abdelhak~M.~Zoubir,~\IEEEmembership{Fellow,~IEEE}
\thanks{M.~Fau\ss{} and A.~M.~Zoubir are with the Signal Processing Group, Institute for Telecommunications, Department of Electrical Engineering and
Information Technology, Technische Universit\"at Darmstadt, Merckstr. 25, 64283 Darmstadt, Germany. E-mail: \{mfauss,zoubir\}@spg.tu-darmstadt.de.}%
\thanks{This work was performed within the LOEWE Priority Program Cocoon (www.cocoon.tu-darmstadt.de) supported by the LOEWE research initiative
of the state of Hesse/Germany.}}

\maketitle

\copyrightnotice

\begin{abstract}
  The density band model proposed by Kassam \cite{Kassam1981} for robust hypothesis testing is revisited. First, a novel criterion for the general
  characterization of least favorable distributions is proposed that unifies existing results. This criterion is then used to derive an implicit
  definition of the least favorable distributions under band uncertainties. In contrast to the existing solution, it only requires two scalar values
  to be determined and eliminates the need for case-by-case statements. Based on this definition, a generic fixed-point algorithm is proposed that
  iteratively calculates the least favorable distributions for arbitrary band specifications. Finally, three different types of robust tests that
  emerge from band models are discussed and a numerical example is presented to illustrate their potential use in practice.
\end{abstract}


%
\IEEEpeerreviewmaketitle

\section{Introduction}

%
%
%
%
\IEEEPARstart{S}{tatistical} hypothesis tests are referred to as \emph{robust}, if they are insensitive to small, random deviations from the
underlying model. In this paper, we consider robustness against \emph{distributional uncertainties}, meaning that, under either hypothesis, the
distribution of the observed random variable is only known approximately. Each hypothesis is hence composite, i.e., it is represented by a set or
class of possible distributions. A test is further called \emph{minimax} robust, if it guarantees a certain maximum error probability over the entire
set of distributions specified by the composite hypotheses. Because of this property, minimax robust tests are often essential for the design of
systems that have to function reliably in harsh environments or cannot be modeled accurately.

The field of robust statistics, and robust hypothesis testing in particular, was developed foremost by Huber in the mid-1960s
\cite{Huber1964,Huber1965}. He was the first to derive the famous clipped likelihood ratio test, which is robust against outliers of the
$\varepsilon$-con\-ta\-mi\-na\-tion type, i.e., infrequent, grossly corrupted observations. This kind of contamination is particularly critical since 
a single corrupted observation can be enough to alter the outcome of a non-robust test \cite{Huber1981}. The clipped likelihood ratio test was further
shown to be a test for two simple hypotheses. More precisely, it is a regular likelihood ratio test of the so-called \emph{least favorable} instead of
the \emph{nominal} distributions, the latter denoting the distributions of the uncontaminated data.

Despite their wide use in practice, $\varepsilon$-con\-ta\-mi\-na\-tion models are often not sufficient for accurately describing the uncertainty in
the distributions. On the one hand, the assumption that the majority of the data follows the nominal distribution exactly can be too optimistic. On
the other hand, the assumption that the outliers are drawn from an arbitrary distribution can be too pessimistic. In addition, many types of
uncertainties cannot easily be incorporated into an $\varepsilon$-con\-ta\-mi\-na\-tion model. Approximately known shapes or positions of
distributions, for example, are usually hard to formulate in terms of nominals and outliers. In such cases, techniques involving the estimation of the
true model may be preferable, like adaptive nonlinearities \cite{Al-Sayed2013} or generalized likelihood ratio tests \cite{Zeitouni1992}. However, in
contrast to minimax solutions, such methods do not guarantee pre-specified error probabilities and require changes in the distributions to happen
slowly enough for the estimates to be updated. For these reasons, more flexible uncertainty models for minimax robust tests are a topic of ongoing
research \cite{Levy2009,Gul2013,Nikolaidis2011}.

In 1981, Kassam published a paper titled ``Robust Hypothesis Testing for Bounded Classes of Probability Densities'' \cite{Kassam1981} in
which he proposed an uncertainty model for probability distributions that later became known as the \emph{band model}. It allows each hypothesis to be
formulated in terms of a density band within which the true density is supposed to lie and generalizes the outlier models suggested by Huber
\cite{Huber1965}, \"Osterreicher \cite{Oesterreicher1978}, and Levy \cite{Levy2008}. It can further be interpreted as both an
$\varepsilon$-contamination model with bounded outlier distributions, or a model  for general uncertainties in the shape that can be specified without
introducing nominals. A more detailed discussion is deferred to later sections.

Considering its generality and versatility, it is astonishing that the band model has received very little attention in the robust statistics
literature. While Huber's seminal paper on $\varepsilon$-contaminated observations has enjoyed an unbroken stream of citations since its publication
in 1965, Kassam's paper, although covering a more general case, did not have a comparable impact. One of the reasons for the limited interest in
\cite{Kassam1981} might be the form of its main result. The theorem stating the least favorable densities spans the space of a column and
distinguishes between four special cases, each involving a piecewise definition of the densities. In order to know which case holds, one has to check
the existence or non-existence of in total six constants that have to be chosen such that the solutions are valid densities. In some cases the
solution involves a function that ``can be found'', but is not specified explicitly. Even though none of these issues is critical, the solution and
its calculation appear to be cumbersome and inelegant, especially compared to the lean results for the $\varepsilon$-contamination model. Many authors
seem to have come to the conclusion that the additional complexity the band model introduces to the least favorable densities is too high a price to
pay for the increase in generality.

As a consequence, the popularity of the band model in robust hypothesis testing has been steadily decreasing. While it used to be one of the standard
models in robust testing and filtering in the 1980s \cite{KassamPoor1985,Poor1983}, today many signal processing practitioners are not aware of
its existence and some books on robust statistics ignore it entirely \cite{Maronna2006,Levy2008}.

In this paper, we revisit the band model, give a detailed description of its properties and motivate its use in practice. Beyond that, our original
contribution is threefold: First, we state a novel criterion for the characterization of least favorable distributions. In our view, this
criterion is conceptually simpler than the existing ones and results in more tractable optimization problems. Second, we derive a concise implicit
definition of the least favorable distributions for the band model that offers additional insight into their structure. Third, based on this
definition, we propose a fixed-point algorithm that provides a simple generic alternative for calculating the densities of the least favorable
distributions without the need for a case analysis.

The paper is organized as follows: In Section~\ref{sec:fundamentals}, we give a brief overview of some fundamental concepts in minimax robust
hypothesis testing. The question of how to characterize least favorable distributions in general is discussed in Section~\ref{sec:characterization}.
After introducing the band model in Section~\ref{sec:model}, we derive its least favorable distributions in Section~\ref{sec:lfds}. In
Section~\ref{sec:computation}, a fixed-point algorithm for their calculation is stated. A detailed discussion of the different types of robust tests
that result from band models is given in Section~\ref{sec:discussion}. In Section \ref{sec:example}, we present an example for how the band model can
be used in practice.

\emph{Notation}: We denote random variables with upper case letters and their realizations with the corresponding lower case letters. Similarly,
probability measures (distributions) are denoted by upper case letters, the corresponding densities by lower case letters. The notation $\{X = x\}$ is
shorthand for $\{\omega \in \Omega: X(\omega) = x\}$ and $P[X = x]$ for $P(\{X = x\})$. The expected value of a random variable $X$ with respect to a
measure $P$ is written as $E_P[X]$. Occasionally, a pair $(x_0,x_1)$ is referred to simply as $x$. This will be clear from the context.

\section{Fundamentals of Minimax Robust Detection}
\label{sec:fundamentals}

Let $(X_1,\ldots,X_n)$ be a sequence of independently and identically distributed random variables with common distribution $P$ defined on some
measurable space $(\Omega,\mathcal{F})$. Throughout the paper, we assume that all probability measures have a continuous density function with respect
to some common measure $\mu$, i.e.,
\begin{equation*}
  \int_B dP = \int_B p \, \dint \mu, \quad \forall B \in \mathcal{F}.
\end{equation*}
We denote the set of all distributions on $(\Omega,\mathcal{F})$ that admit this property by $\Pmu$.

The goal of a binary statistical hypothesis test is to decide between the two hypotheses
\begin{align*}
  \mathcal{H}_0\colon & \quad P = P_0, \\
  \mathcal{H}_1\colon & \quad P = P_1,
\end{align*}
where $P_0, P_1 \in \Pmu$ are two given distributions and $\mathcal{H}_0$ and $\mathcal{H}_1$ are referred to as the null and alternative hypothesis,
respectively. A statistical test for $\mathcal{H}_0$ against $\mathcal{H}_1$ is in general defined by a decision $d \in \{0,1\}$ and a (randomized)
decision rule
\begin{equation*}
  \delta\colon \Omega^n \to [0,1],
\end{equation*}
where $\delta = \delta(x_1,\ldots,x_n)$ denotes the conditional probability to decide for the alternative hypothesis, given the observations
$(x_1,\ldots,x_n)$. The set of all decision rules is denoted by $\Delta$. The type I and type II error probabilities are given by
\begin{align*}
  P_0[d=1] &= E_{P_0}[\,\delta\,], \\
  P_1[d=0] &= E_{P_1}[1-\delta].
\end{align*}
The optimal decision rule $\delta^*$ for the simple binary hypothesis test is a threshold comparison of the likelihood ratio, i.e.,
\begin{equation}
  \label{eq:lr_test}
  \delta^* = \begin{cases}
                1,      & z_n > \eta \\
                \kappa, & z_n = \eta \\
                0,      & z_n < \eta,
             \end{cases}
\end{equation}
where $\eta > 0$ is the threshold value, $\kappa \in [0,1]$ can be chosen arbitrarily, and $z_n\colon \Omega^n \to \mathbb{R}_+$ denotes the
likelihood ratio
\begin{equation*}
  z_n \vcentcolon= \prod_{k=1}^n\frac{\dint P_1}{\dint P_0}(x_k) = \prod_{k=1}^n\frac{p_1(x_k)}{p_0(x_k)}.
\end{equation*}
The likelihood ratio test is optimal in a very general sense \cite{Lehmann2005}. In particular, it minimizes the weighted sum error
probability, i.e., it solves
\begin{equation}
  \label{eq:decision_rule_simple}
  \min_{\delta \in \Delta} \; E_{P_0}[\,\delta\,] + \lambda \, E_{P_1}[1-\delta],
\end{equation}
where the weighting factor $\lambda = 1/\eta \geq 0$ determines the threshold value in \eqref{eq:lr_test}. The robust version of
\eqref{eq:decision_rule_simple} is considered in the following sections. However, we want to emphasize that the resulting minimax solution is optimal
also in a Neyman--Pearson and a Bayesian sense \cite{Huber1965}.

In robust testing, the distribution under each hypothesis is assumed not to be known exactly. This distributional uncertainty is modeled by two
disjoint sets $\mathcal{P}_0,\mathcal{P}_1 \subset \Pmu$ so that the hypotheses become
\begin{equation}
  \begin{aligned}
    \mathcal{H}_0\colon & \quad P \in \mathcal{P}_0, \\
    \mathcal{H}_1\colon & \quad P \in \mathcal{P}_1.
  \end{aligned} \label{eq:uncertainty_model_iid}
\end{equation}
It is worth noting that this model can be relaxed to
\begin{equation}
  \begin{aligned}
    \mathcal{H}_0\colon & \quad P_k \in \mathcal{P}_0, \\
    \mathcal{H}_1\colon & \quad P_k \in \mathcal{P}_1.
  \end{aligned}\label{eq:uncertainty_model_i}
\end{equation}
for all $k \in \{1,\ldots,n\}$, where $P_k$ denotes the distribution of $X_k$. That is, $(X_1,\ldots,X_n)$ can be assumed to be a sequence of
independent, but not necessarily identically distributed random variables. Both models result in the same minimax robust test
\cite{Huber1965,Huber1973}. For the sake of a more compact notation, the composite hypotheses are assumed to be of the form
\eqref{eq:uncertainty_model_iid} throughout the paper. However, the results also apply to the more general case \eqref{eq:uncertainty_model_i}.

Tests for sets of distributions are known as composite hypothesis tests and have been studied extensively in the literature
\cite{Papoulis1991,Lehmann2005}. What distinguishes minimax robust procedures from other approaches is that the test is designed \emph{a priori} so as
to guarantee a certain reliability for all possible pairs $(P_0,P_1) \in \mathcal{P}_0 \times \mathcal{P}_1$. Mathematically, this property is
formulated in terms of a minimax problem. The robust testing problem corresponding to \eqref{eq:decision_rule_simple} is thus given by
\begin{equation}
  \label{eq:decision_rule_robust}
  \min_{\delta \in \Delta} \; \max_{(P_0,P_1) \in \mathcal{P}_0 \times \mathcal{P}_1} \; E_{P_0}[\,\delta\,] + \lambda \, E_{P_1}[1-\delta].
\end{equation}

A decision rule and a pair of distributions that solve \eqref{eq:decision_rule_robust} are called minimax optimal. The existence and
characterization of minimax optimal solutions, however, is an intricate question. It has received continuous attention in applied mathematics,
physics, and economics since the 1950s and is still an active area of research \cite{Sion1958,Simons1995,Seung2007}.

The most useful minimax theorem in the context of robust hypothesis testing is due to Sion \cite{Komiya1988}. In a nutshell, it follows from Sion's
minimax theorem that a minimax optimal test exists for compact convex sets $\mathcal{P}_0, \mathcal{P}_1$ and is again a likelihood ratio test of the
form \eqref{eq:lr_test}, but with the nominal distributions $P_0$ and $P_1$ replaced by the least favorable distributions $Q_0$ and $Q_1$. The
design of minimax robust tests hence reduces to finding the pair $(Q_0,Q_1)$. This, however, is a non-trivial problem in itself. In the next section,
we review two commonly used characterizations for least favorable distributions and propose a third one that we find to be simpler and more suitable
in an optimization  context.

\section{Criteria for Least Favorable Distributions}
\label{sec:characterization}

The first criterion for least favorable distributions was given by Huber \cite{Huber1964}, who used it to prove minimax optimality of the clipped
likelihood ratio test. It is now known by the name \emph{stochastic dominance}.

\begin{theorem}[Stochastic Dominance]
  \label{th:lfds_characterization_1}
  A pair of distributions $(Q_0,Q_1) \in \mathcal{P}_0 \times \mathcal{P}_1$ is least favorable for all sample sizes $n \geq 1$, i.e., minimax
  optimal in combination with a likelihood ratio test of the form \eqref{eq:lr_test}, if it fulfills
  \begin{equation}
    \label{eq:lfds_characterization_1}
    \begin{aligned}
      Q_0 \left[ \frac{q_1}{q_0} >    \eta \right] & \geq P_0\left[ \frac{q_1}{q_0} >    \eta \right] \quad \text{and} \\
      Q_1 \left[ \frac{q_1}{q_0} \leq \eta \right] & \geq P_1\left[ \frac{q_1}{q_0} \leq \eta \right]
    \end{aligned}
  \end{equation}
  for all $(P_0,P_1) \in \mathcal{P}_0 \times \mathcal{P}_1$ and all $\eta \geq 0$.
\end{theorem}

The interpretation of Theorem~\ref{th:lfds_characterization_1} is that the distributions $Q_0$ and $Q_1$ have to be chosen such that for a single
sample test the probabilities of both error types are jointly maximized, irrespective of the value of the likelihood ratio threshold. Stochastic
dominance is a natural and widely used criterion for minimax optimality \cite{Gul2013,Veeravalli1994}.

About a decade later, Huber and Strassen derived a different, more technical criterion to characterize least favorable distributions.
\begin{theorem}[Theorem 6.1 in \cite{Huber1973}]
  \label{th:lfds_characterization_2}
  Let $\Psi \vcentcolon [0,1] \to \mathbb{R}$ be a convex function that is twice continuously differentiable. A pair of distributions $(Q_0,Q_1)
  \in \mathcal{P}_0 \times \mathcal{P}_1$ is least favorable in the sense of Theorem \ref{th:lfds_characterization_1} if it  minimizes
  \begin{equation}
    H_{\Psi}(P_0,P_1) = \int \Psi\left( \frac{p_0}{p_0+p_1} \right) \, \dint(P_0+P_1) \label{eq:lfds_characterization_2}
  \end{equation}
  for all $\Psi$ and among all $(P_0,P_1) \in \mathcal{P}_0 \times \mathcal{P}_1$.
\end{theorem}

Theorem \ref{th:lfds_characterization_2} states that the least favorable distributions concurrently minimize all $f$-divergences induced by $\Psi$
among $(P_0,P_1) \in \mathcal{P}_0 \times \mathcal{P}_1$.

Qualitatively speaking, the idea underlying stochastic dominance is that the least favorable distributions should maximize the error probabilities,
while Theorem \ref{th:lfds_characterization_2} corresponds to the more abstract intuition that the least favorable distributions should be ``as
similar as possible''.

We propose a criterion that lends itself to both interpretations and, as we believe, helps to simplify and unify the theory of robust testing.
\begin{theorem}
  \label{th:lfds_characterization_3}
  A pair of distributions $(Q_0,Q_1) \in \mathcal{P}_0  \times \mathcal{P}_1$ is least favorable in the sense of Theorem
  \ref{th:lfds_characterization_1}, if it maximizes
  \begin{equation}
    \label{eq:lfds_characterization_3}
    L_{\lambda}(P_0,P_1) = \int \min\{ p_0, \lambda \, p_1 \} \, \dint \mu
  \end{equation}
  for all $\lambda \geq 0$ and among all $(P_0,P_1) \in \mathcal{P}_0 \times \mathcal{P}_1$.
\end{theorem}

A proof is detailed in Appendix~\ref{apx:proof_lfds_characterization}. The two aspects of error maximization and distance minimization are both
captured in \eqref{eq:lfds_characterization_3}. For example, evaluating \eqref{eq:decision_rule_simple} for $n=1$ yields
\begin{equation*}
  \begin{split}
    \min_{\delta \in \Delta} \, E_{P_0}[\delta] + \lambda E_{P_1}[1-\delta] &= \min_{\delta \in \Delta} \int \delta p_0 + \lambda (1-\delta) p_1 \,
      \dint \mu \\
    &= \int \min\{ p_0, \lambda p_1 \} \, \dint \mu,
  \end{split}
\end{equation*}
which is the expression given in Theorem~\ref{th:lfds_characterization_3}. This means that the least favorable distributions concurrently maximize the
weighted error sum of a single sample test for all weighting coefficients $\lambda \geq 0$, which is in close analogy to the stochastic dominance
characterization. At the same time, it can be shown that maximizing \eqref{eq:lfds_characterization_3} is equivalent to maximizing
\begin{equation*}
  \int \min \left\{ \frac{p_0}{p_0+p_1} , \frac{\lambda}{1+\lambda} \right\} \, \dint (P_0+P_1),
\end{equation*}
which, disregarding the differentiability assumption, is a special case of Theorem \ref{th:lfds_characterization_2} with $\Psi$ chosen to be $\Psi(s)
= -\min\{s,\lambda/(1+\lambda)\}$.

Before proceeding further, we would like to discuss briefly the relations between the three theorems stated above. In \cite{Huber1973}, it is shown
that Theorem~\ref{th:lfds_characterization_2} implies Theorem~\ref{th:lfds_characterization_1}. Theorem~\ref{th:lfds_characterization_3} in turn can
be shown to imply Theorem~\ref{th:lfds_characterization_2}. This is the case since a pair of distributions that satisfies
Theorem~\ref{th:lfds_characterization_3} minimizes all $f$-divergences, whereas a pair of distributions that satisfies
Theorem~\ref{th:lfds_characterization_2} only minimizes $f$-divergences for which $\Psi$ is twice differentiable. Moreover, all of the three theorems
are sufficient, but none of them necessary. For single sample tests with a given likelihood ratio threshold, for example,
Theorem~\ref{th:lfds_characterization_1} is sufficient, but not necessary because it requires $Q_0$ and $Q_1$ to be least favorable for \emph{all}
thresholds. Consequently, Theorem~\ref{th:lfds_characterization_2} and Theorem~\ref{th:lfds_characterization_3} are not necessary either.

\section{The Band Model}
\label{sec:model}

The band model proposed by Kassam in \cite{Kassam1981} covers composite hypotheses specified by sets of the form
\begin{equation}
  \label{eq:density_corridor}
  \Pcor = \{ P \in \Pmu \,:\, \pmin \leq p \leq \pmax \},
\end{equation}
where $\pmin$ and $\pmax$ fulfill
\begin{equation*}
  0 \leq \pmin \leq \pmax, \quad \Pmin(\Omega) \leq 1, \quad \Pmax(\Omega) \geq 1.
\end{equation*}
In words, it restricts the true density to lie within a band specified by $\pmin$ and $\pmax$. This is indicated by the notation
$\Pcor$. Note that $\Pmin$ and $\Pmax$ are measures on $(\Omega,\mathcal{F})$, but usually not probability measures, and $\Pmax$ does not need
to be finite.

Alternatively, the band model can be interpreted as an $\varepsilon$-contamination model with bounded outlier distribution. In order to see this,
note that every $p$, with $P \in \Pcor$, can be written as $p = \pmin + \varepsilon h$, where
\begin{equation}
  \label{eq:def_gamma}
  \varepsilon = 1-\Pmin(\Omega)
\end{equation}
and $h$ denotes the density of an outlier distribution $H \in \Pmu$. In contrast to the $\varepsilon$-contamination model, however, not every $H$ is
feasible under the density band model. More precisely, $h$ has to be chosen such that
\begin{equation*}
  \pmin \leq \pmin + \varepsilon h \leq \pmax
\end{equation*}
which yields the constraint
\begin{equation}
  \varepsilon h \leq \pmax-\pmin.
  \label{eq:h_constraint}
\end{equation}
By definition of $\Pmu$, both sides of \eqref{eq:h_constraint} can be integrated over all $B \in \mathcal{F}$ so that \eqref{eq:h_constraint} can
equivalently be written as
\begin{equation*}
  \varepsilon H \leq \Pmax-\Pmin
\end{equation*}
and \eqref{eq:density_corridor} can be written as
\begin{equation}
  \Pcor = \{P \in \Pmu \,:\, P = \Pmin + \varepsilon H \,,\, \varepsilon H \leq \Pmax-\Pmin \}.
  \label{eq:density_corridor_outlier}
\end{equation}
In this regard, the band model is an $\varepsilon$-contamination model that allows the incorporation of \emph{a priori} knowledge in the form of
additional constraints on the outlier distribution.

\section{Least Favorable Distributions for the Band Model}
\label{sec:lfds}

In this section, we state and discuss the main result of the paper, which is an implicit characterization of the pair of least favorable
distributions for the band model \eqref{eq:density_corridor}.

\subsection{An Implicit Characterization}
\begin{theorem}
  \label{th:band_lfds}
  Given hypotheses of the form \eqref{eq:density_corridor}, the pair $(Q_0,Q_1) \in \Pcor_0 \times \Pcor_1$ is least favorable
  in the sense of Theorem \ref{th:lfds_characterization_1}, if the pair of densities $(q_0,q_1)$ satisfies
  \begin{equation}
    \label{eq:band_lfds}
    \begin{aligned}
      q_0 & = \min \{ \pmax_0 \,,\, \max \{ c_0 (\alpha q_0 + q_1) \,,\, \pmin_0 \} \}, \\
      q_1 & = \min \{ \pmax_1 \,,\, \max \{ c_1 (q_0 + \alpha q_1) \,,\, \pmin_1 \} \}
    \end{aligned}
  \end{equation}
  for some $\alpha \geq 0$ and some $c_0, c_1 \in [0,\tfrac{1}{\alpha}]$. Such a pair always exists.
\end{theorem}

The proof of Theorem \ref{th:band_lfds} is given by first deriving an upper bound on the function $L_{\lambda}$ in Theorem
\ref{th:lfds_characterization_3} and then showing that the densities in \eqref{eq:band_lfds} attain this bound. Their existence is shown in
Section~\ref{sec:computation} by means of a constructive algorithm.
\begin{theorem}
  \label{th:upper_bound}
  Given hypotheses of the form \eqref{eq:density_corridor}, for all $(P_0,P_1) \in \Pcor_0 \times \Pcor_1$ it is the case that $L_{\lambda}$ in
  Theorem \ref{th:lfds_characterization_3} is upper bounded by
  \begin{equation*}
    L_{\lambda}(P_0,P_1) \leq \int \min \{ \hat{q}_0 \,,\, \lambda \, \hat{q}_1 \} \, \dint \mu + v_0 \varepsilon_0 + \lambda v_1 \varepsilon_1,
  \end{equation*}
  where
  \begin{equation*}
    \hat{q}_i = v_i \pmin_i + (1-v_i) \pmax_i, \quad i=0,1,
  \end{equation*}
  $\varepsilon_0, \varepsilon_1 \in [0,1]$ are defined in \eqref{eq:def_gamma} and $v_0,v_1 \in [0,1]$ can be chosen arbitrarily.
\end{theorem}

A proof is laid down in Appendix \ref{apx:proof_upper_bound}. In the following paragraphs, it will become clear that the bound in
Theorem~\ref{th:upper_bound} can be tightened to $v_0, v_1 \in \{0,1\}$. We chose to state it in this more relaxed form to emphasize that an upper
bound on $L_{\lambda}$ can be obtained from every convex combination of the upper and lower bound. 

We now show that for every feasible combination of $\lambda$, $\alpha$ and $(c_0, c_1)$, the densities in \eqref{eq:band_lfds} attain the upper bound
in Theorem \ref{th:upper_bound} for some choice of $v_0$ and $v_1$. Four cases are covered separately, but only two of them in detail since the others
follow analogously.

\emph{Case 1:} $c_0 > \lambda/(1+\alpha\lambda)$ and $c_1 \geq 1/(\alpha+\lambda)$

On the set $\{q_0 \leq \lambda q_1\}$ it holds that
\begin{align}
  q_0 &\leq \lambda q_1 \notag \\
  (1+\alpha\lambda) q_0 &\leq \lambda (\alpha q_0 + q_1) \notag \\
  q_0 &\leq \frac{\lambda}{1+\alpha\lambda } (\alpha q_0 + q_1). \label{eq:case1_01}
\end{align}
Inserting $q_0$ from \eqref{eq:band_lfds} into the left hand side of \eqref{eq:case1_01} yields
\begin{equation*}
  \min \{ \pmax_0 \,,\, \max \{ c_0 (\alpha q_0 + q_1) \,,\, \pmin_0 \} \} \leq \frac{\lambda}{1+\alpha\lambda} (\alpha q_0 + q_1).
\end{equation*}
For $c_0 > \lambda/(1+\alpha\lambda)$ this can only be satisfied if
\begin{equation*}
  \pmax_0 \leq \frac{\lambda}{1+\alpha\lambda} (\alpha q_0 + q_1),
\end{equation*}
which in turn implies
\begin{equation}
  q_0 = \pmax_0 \quad \text{on} \quad \{q_0 \leq \lambda q_1\}. \label{eq:case1_02}
\end{equation}
Analogously, on the set $\{q_0 > \lambda q_1\}$ it holds that
\begin{align}
  q_0 &> \lambda q_1 \notag \\
  q_0 + \alpha q_1 &> (\alpha+\lambda) q_1 \notag \\
  \frac{1}{\alpha+\lambda} (q_0 + \alpha q_1) &> q_1. \label{eq:case1_11}
\end{align}
Inserting $q_1$ from \eqref{eq:band_lfds} into the right hand side of \eqref{eq:case1_11} yields
\begin{equation*}
  \frac{1}{\alpha+\lambda}(q_0 + \alpha q_1) > \min \{ \pmax_1 \,,\, \max \{ c_1 (q_0 + \alpha q_1) \,,\, \pmin_1 \} \}.
\end{equation*}
For $c_1 \geq 1/(\alpha+\lambda)$, this can only be satisfied if
\begin{equation*}
  \frac{1}{\alpha+\lambda}(q_0 + \alpha q_1) > \pmax_1,
\end{equation*}
which in turn implies
\begin{equation}
  q_1 = \pmax_1 \quad \text{on} \quad \{q_0 > \lambda q_1\}. \label{eq:case1_12}
\end{equation}
Combining \eqref{eq:case1_02} and \eqref{eq:case1_12} yields
\begin{equation*}
  \int \min \{ q_0 \,,\, \lambda q_1 \} \, \dint \mu = \int \min \{ \pmax_0 \,,\, \lambda \pmax_1 \} \, \dint \mu,
\end{equation*}
which is the upper bound in Theorem \ref{th:upper_bound} evaluated at $v_0 = v_1 = 0$.

\emph{Case 2:} $c_0 \leq \lambda/(1+\alpha\lambda)$ and $c_1 < 1/(\alpha+\lambda)$

On the set $\{q_0 > \lambda q_1\}$ it holds that
\begin{align}
  q_0 &> \lambda q_1 \notag \\
  (1+\alpha \lambda) q_0 &> \lambda (\alpha q_0 + q_1) \notag \\
  q_0 &> \frac{\lambda}{1+\alpha \lambda} (\alpha q_0 + q_1). \label{eq:case2_01}
\end{align}
Inserting $q_0$ from \eqref{eq:band_lfds} into the left hand side of \eqref{eq:case2_01} yields
\begin{equation*}
  \min \{ \pmax_0 \,,\, \max \{ c_0(\alpha q_0 + q_1) \,,\, \pmin_0 \} \} > \frac{\lambda}{1+\alpha \lambda} (\alpha q_0 + q_1)
\end{equation*}
For $c_0 \leq \lambda/(1 + \alpha \lambda)$ this can only be satisfied if
\begin{equation}
  q_0 = \pmin_0 \quad \text{on} \quad \{q_0 > \lambda q_1\}. \label{eq:case2_02}
\end{equation}
That is, the least favorable distribution $q_0$ equals its lower bound on $\{q_0 > \lambda q_1\}$. As a consequence, the outlier
distribution $H_0$, which satisfies $Q_0 = \Pmin_0 + \varepsilon_0 H_0$, is concentrated on $\{q_0 \leq \lambda q_1\}$.

Analogously, on the set $\{q_0 \leq \lambda q_1\}$ it holds that
\begin{align}
  q_0 &\leq \lambda q_1 \notag \\
  q_0 + \alpha q_1 &\leq (\alpha+\lambda) q_1 \notag \\
  \frac{1}{\alpha+\lambda} (q_0 + \alpha q_1) &\leq q_1. \label{eq:case2_11}
\end{align}
Inserting $q_1$ from \eqref{eq:band_lfds} into the right hand side of \eqref{eq:case2_11} yields
\begin{align*}
  \frac{1}{\alpha+\lambda}(q_0 + \alpha q_1) \leq \min \{ \pmax_1 \,,\, \max \{ c_1(q_0 + \alpha q_1) \,,\, \pmin_1 \} \}
\end{align*}
For $c_1 < 1/(\alpha+\lambda)$ this can only be satisfied if
\begin{equation}
  \label{eq:case2_12}
  q_1 = \pmin_1 \quad \text{on} \quad \{q_0 \leq \lambda q_1\}.
\end{equation}
Consequently, $H_1$ is concentrated on $\{q_0 > \lambda q_1\}$. Combining \eqref{eq:case2_12} and \eqref{eq:case2_02} yields
\begin{equation*}
  \begin{split}
    \int \min \{ q_0, \lambda q_1\} \, \dint \mu &= \int \min \{ \pmin_0 + \varepsilon_0 h_0 \,,\, \lambda (\pmin_1 + \varepsilon_1 h_1) \} \,
      \dint \mu \\
    &= \int \min \{ \pmin_0 \,,\, \lambda \pmin_1 \} \, \dint \mu + \varepsilon_0 + \lambda \varepsilon_1,
  \end{split}
\end{equation*}
which is the upper bound in Theorem \ref{th:upper_bound} evaluated at $v_0 = v_1 = 1$.

\emph{Case 3:} $c_0 > \lambda/(1+\alpha\lambda)$ and $c_1 < 1/(\alpha+\lambda)$

Combining the arguments from the previous cases, it can be shown that
\begin{equation*}
  q_0 = \pmax_0 \quad \text{on} \quad \{q_0 \leq \lambda q_1\},
\end{equation*}
and
\begin{equation*}
  q_1 = \pmin_1 \quad \text{on} \quad \{q_0 \leq \lambda q_1\},
\end{equation*}
which corresponds to $v_0 = 0$ and $v_1 = 1$ in Theorem \ref{th:upper_bound}.

\emph{Case 4:} $c_0 \leq \lambda/(1+\alpha\lambda)$ and $c_1 \geq 1/(\alpha+\lambda)$

Again, using the same arguments as before, we obtain
\begin{equation*}
  q_0 = \pmin_0 \quad \text{on} \quad \{q_0 > \lambda q_1\},
\end{equation*}
and
\begin{equation*}
  q_1 = \pmax_1 \quad \text{on} \quad \{q_0 > \lambda q_1\},
\end{equation*}
which corresponds to $v_0 = 1$ and $v_1 = 0$ in Theorem \ref{th:upper_bound}. \hfill $\Square$

\subsection{Discussion}
First, we point out two interesting special cases of Theorem~\ref{th:band_lfds}. Choosing $\alpha=0$ yields
\begin{equation}
  \label{eq:band_lfds_0}
  \begin{aligned}
    q_0 & = \min \{ \pmax_0 \,,\, \max \{ c_0 q_1 \,,\, \pmin_0 \} \}, \\
    q_1 & = \min \{ \pmax_1 \,,\, \max \{ c_1 q_0 \,,\, \pmin_1 \} \}.
  \end{aligned}
\end{equation}
This is likely to be the most intuitive form of Theorem~\ref{th:band_lfds} and the most useful in practice---see Section~\ref{sec:computation}. It
closely resembles the structure of the asymptotic minimax solution $\tilde{Q}_0$, $\tilde{Q}_1$ derived in \cite{Dabak1993}, which for general
uncertainty sets $\mathcal{P}_0$ and $\mathcal{P}_1$ is given by
\begin{equation}
  \label{eq:asymp_lfds}
  \begin{aligned}
    \tilde{Q}_0 & = \min_{P_0 \in \mathcal{P}_0} \D{\text{KL}}{P_0}{\tilde{Q}_1}, \\
    \tilde{Q}_1 & = \min_{P_1 \in \mathcal{P}_1} \D{\text{KL}}{\tilde{Q}_0}{P_1},
  \end{aligned}
\end{equation}
where $D_{\text{KL}}$ denotes the Kullback--Leibler (KL) divergence. In \eqref{eq:asymp_lfds}, $\tilde{Q}_0$ ($\tilde{Q}_1$) is the projection of
$\tilde{Q}_1$ ($\tilde{Q}_0$) onto $\mathcal{P}_0$ ($\mathcal{P}_1$) with respect to the KL divergence. In \eqref{eq:band_lfds_0} this projection
is performed with respect to general $f$-divergences and the projection operator is written out explicitly. Note that asymptotically least favorable
distributions exist for every convex uncertainty set, while the existence of strictly least favorable distributions depends on the uncertainty set.

A problem that arises when stating Theorem~\ref{th:band_lfds} with $\alpha = 0$ is that \eqref{eq:band_lfds_0} is a sufficient but not a necessary
condition for $Q_0$ and $Q_1$ to be least favorable. Assume, for example, that the two density bands do not overlap, i.e., $\{\pmax_0 > 0\}
\cap \{\pmax_1 > 0\} = \emptyset$. In this case, $c_0$ and $c_1$ can be chosen arbitrarily large without ever producing valid densities on the right
hand side of \eqref{eq:band_lfds_0}. Nevertheless, every feasible pair is least favorable in the sense of Theorem~\ref{th:lfds_characterization_1}. By
adding an auxiliary density with sufficient overlap on the right hand side of \eqref{eq:band_lfds_0}, this problem can be avoided. Choosing this
auxiliary density as the least favorable density itself guarantees that the overlap is indeed sufficient and that optimality still holds.

In general, choosing $\alpha > 0$ guarantees that $q_0$ and $q_1$ in \eqref{eq:band_lfds} are well defined. However, a second noteworthy special case
is $\alpha=1$, for which Theorem~\ref{th:band_lfds} becomes
\begin{align*}
  q_0 &= \min \{ \pmax_0 \,,\, \max \{ c_0 (q_0+q_1) \,,\, \pmin_0 \} \}, \\
  q_1 &= \min \{ \pmax_1 \,,\, \max \{ c_1 (q_0+q_1) \,,\, \pmin_1 \} \}.
\end{align*}
In this form, $Q_0$ and $Q_1$ are the projections of the distribution $\tfrac{1}{2}(Q_0+Q_1)$ onto $\Pcor_0$ and $\Pcor_1$, respectively. Since these
projections are unique, knowledge of $\tfrac{1}{2}(Q_0+Q_1)$ is sufficient to obtain $Q_0$ and $Q_1$. In this sense, there exists a \emph{single}
least favorable distribution whose projections onto the respective bands form the least favorable \emph{pair}. In applications, this property might be
used to trade memory for computing power by storing only $\tfrac{1}{2}(Q_0+Q_1)$ and calculating $Q_0$ and $Q_1$ on demand.

We also point out that irrespective of the choice for $\alpha$ in Theorem~\ref{th:band_lfds}, the two constants $c_0$, $c_1$ are sufficient for
characterizing the solution. On the one hand, this is a major simplification compared to the six constants introduced in \cite{Kassam1981}. On the
other hand, it is in perfect analogy to the $\varepsilon$-contamination model and shows how closely the two models are related.

Let us investigate this relation a bit further. The likelihood ratio of the densities in Theorem~\ref{th:band_lfds} can take on six possible values,
namely,
\begin{equation}
  \frac{q_1}{q_0} \in \left\{ \frac{\pmin_1}{\pmin_0} \,,\, \frac{\pmax_1}{\pmin_0} \,,\, \frac{\pmin_1}{\pmax_0} \,,\, \frac{\pmax_1}{\pmax_0} \,,\,
  \frac{1-\alpha c_0}{c_0} \,,\, \frac{c_1}{1-\alpha c_1} \right\}. \label{eq:likelihood_cases}
\end{equation}
Note that some of the terms can be zero or involve a division by zero, which corresponds to observations that are possible only under one of the
hypotheses and hence lead to an unambiguous decision against the impossible hypothesis. The first four terms in \eqref{eq:likelihood_cases} are
obtained by considering only cases where both $q_0$ and $q_1$ are equal to either the lower or the upper bound of the density band. The two remaining
terms are obtained as follows: Assuming that $p_1' < q_1 < p_1''$ holds, we have
\begin{equation*}
  q_1 = c_1(q_0 + \alpha q_1).
\end{equation*}
Dividing both sides by $q_0$ yields
\begin{equation*}
  \frac{q_1}{q_0} = c_1\left(1 + \alpha \frac{q_1}{q_0}\right).
\end{equation*}
Solving this expression for $q_1/q_0$ yields
\begin{equation*}
  \frac{q_1}{q_0} = \frac{c_1}{1 - \alpha c_1}.
\end{equation*}
This result holds irrespective of how $q_0$ is chosen, meaning that we do not need to distinguish between the three cases $q_0 = p_0'$, $q_0 = p_0''$
and $p_0' < q_0 < p_0''$. Analogously, for $p_1' < q_1 < p_1''$ it holds that
\begin{equation*}
  q_0 = c_0(\alpha q_0 + q_1),
\end{equation*}
which when divided by $q_1$ yields
\begin{equation*}
  \frac{q_1}{q_0} = \frac{1-\alpha c_0}{c_0}.
\end{equation*}
This result is independent of $q_1$.

Letting $\pmax_0, \pmax_1 \to \infty$ and setting $\alpha = 0$, we obtain Huber's least favorable densities for the $\varepsilon$-contamination
model\footnote{The case-by-case definition given by Huber is more common in the literature. It can be shown that both definitions are
identical.}
\begin{equation*}
  \begin{aligned}
    q_0 & = \max \{ c_0 q_1 \,,\, \pmin_0 \}, \\
    q_1 & = \max \{ c_1 q_0 \,,\, \pmin_1 \},
  \end{aligned}
\end{equation*}
with the corresponding clipped likelihood ratio
\begin{equation*}
  \frac{q_1}{q_0} \in \left\{ \frac{\pmin_1}{\pmin_0} \,,\, \frac{1}{c_0} \,,\, c_1 \right\}.
\end{equation*}

These basic derivations show how the $\varepsilon$-contamination model emerges naturally as a special case of the band model, which is in contrast to
the prevailing perception that density bands are a rather tedious generalization of $\varepsilon$-contamination.

\section{Computation of the Least Favorable Densities}
\label{sec:computation}

In this section, we propose a fixed-point algorithm that makes use of the implicit definition in Theorem \ref{th:band_lfds} to successively
approximate $q_0$ and $q_1$. It offers a generic, conceptually simple, and easy-to-implement way to determine the least favorable densities without
the need for a case analysis.

First, let the functions $g_i: \mathbb{R} \to \mathbb{R}_+$ be defined by
\begin{equation*}
  g_i(c \,;\, p) \vcentcolon= \int \min \{ \pmax_i \,,\, \max \{ c \, p \,,\, \pmin_i \} \} \, \dint \mu - 1
\end{equation*}
where $i=0,1$ and $p > 0$ is any $\mu$-integrable function.
\begin{lemma}
  \label{lm:root_f}
  Given any pair of distributions $P_0 \in \Pcor_0$, $P_1 \in \Pcor_1$, the functions $g_0(c_0 \,;\, \alpha p_0+p_1), g_1(c_1
  \,;\, p_0+\alpha p_1)$ are nondecreasing and continuous and there exist some $c_0^*, c_1^* \in [0,\tfrac{1}{\alpha}]$ such that $g_0(c_0^* \,;
  \alpha p_0+p_1) = 0$ and  $g_1(c_1^* \,;  p_0+\alpha p_1) = 0$.
\end{lemma}

A proof is detailed in Appendix \ref{apx:proof_root_f}. The algorithm we propose is given in Table~\ref{alg:fixedpoint}.
\begin{table}
  \centering
  \begin{framed}
    \begin{algorithmic}[1]
      \STATE \textbf{input:} $\pmin_0$, $\pmax_0$, $\pmin_1$, $\pmax_1$, $\alpha \geq 0$
      \STATE \textbf{initialize:} Choose feasible initial densities $q_0^0 \in \Pcor_0$, $q_1^0 \in \Pcor_1$ and set $n \leftarrow 0$.
      \REPEAT
        \STATE Set $n \leftarrow n+1$
        \STATE Solve $g_0(c_0 \,; \alpha q_0^{n-1}+q_1^{n-1}) = 0$ for $c_0$ and set
               \begin{equation*}
                 q_0^n \leftarrow \min \{ \pmax_0 \,,\, \max \{ c_0(\alpha q_0^{n-1}+q_1^{n-1}) \,,\, \pmin_0 \} \}.
               \end{equation*}
        \STATE Solve $g_1(c_1 \,; q_0^n+\alpha q_1^{n-1}) = 0$ for $c_1$ and set
               \begin{equation*}
                 q_1^n \leftarrow \min \{ \pmax_1 \,,\, \max \{ c_1 (q_0^n+\alpha q_1^{n-1}) \,,\, \pmin_1 \} \}.
               \end{equation*}
        
      \UNTIL {$q_0^n \approx q_0^{n-1}$ and $q_1^n \approx q_1^{n-1}$}
      \RETURN $q_0^n$, $q_1^n$
    \end{algorithmic}
  \end{framed}
  \label{alg:fixedpoint}
  \caption{Iterative algorithm to calculate the least favorable densities for the band model \eqref{eq:density_corridor}.}
\end{table}
It reduces the problem of determining the least favorable densities to a repeated search for the root of a monotonic and continuous function.
Convergence of Algorithm 1 is proven in Appendix \ref{apx:proof_convergence_alg1}.

The least favorable densities $q_0$ and $q_1$ are not necessarily unique, cf.~\cite{Kassam1981}. Therefore, the solution of Algorithm 1 depends
on the initial densities $q_0^0$ and $q_1^0$. The minimax optimality of the robust test is not affected by this dependence.

The termination criterion in line seven is intentionally left vague. One option is to require that $\max_{i \in \{0,1\}} \lVert q_i^n-q_i^{n-1}
\rVert$ is sufficiently small, where $\lVert \cdot \rVert$ denotes a suitable norm. Alternatively, some $f$-divergence between $Q_0^n$ and $Q_1^n$
or the likelihood ratio can be tracked.

The speed of convergence depends on the choice of $\alpha$. By inspection of the update rule of the iterative algorithm, it can be seen that the
updated density, say $q_0^n$, is the projection of a weighted sum of the previous densities, i.e., $\alpha q_0^{n-1} + q_1^{n-1}$. Increasing $\alpha$
increases the influence of the previous iterate $q_0^{n-1}$ on the next next iterate $q_0^n$. Hence, the ``step size'' of the algorithm depends on
$\alpha$ and it converges slower for large $\alpha$. Therefore, $\alpha$ should be chosen as small as possible to achieve fast convergence and in most
cases $\alpha = 0$ is the best option. Nonzero $\alpha$-values are necessary only when the joint support of the least favorable distributions
is very small. The easiest way to identify such cases is running Algorithm 1 with $\alpha = 0$ and checking whether or not a solution for $c_0$ and
$c_1$ exists.

In general, the proposed algorithm is stable and does not pose any numerical challenges. The convergence is guaranteed to be monotonic in all
$f$-divergences for $\alpha = 0$ and monotonic in the total variation distance for $\alpha > 0$, see also Appendix~\ref{apx:proof_convergence_alg1}.
The computation time depends foremost on how costly it is to evaluate the integral in the definition of $g$.

\section{Robust Tests Resulting from Band Models}
\label{sec:discussion}

Statements can be found in the literature which claim that ``the robust solution for the band model [\ldots] is similar to that for the
$\varepsilon$-contamination model'' \cite{KassamPoor1985}. In this section we show that this is not true in general. More precisely, we identify three
characteristic types of robust tests that a band model can produce. These are explained and illustrated using a
simple Gaussian example with lower bounds
\begin{equation}
  \begin{aligned}
    \pmin_0(x) & = 0.8 \, p_{\mathcal{N}}(x \,;\, -1,2), \\
    \pmin_1(x) & = 0.8 \, p_{\mathcal{N}}(x \,;\, 1,2),
  \end{aligned}
\label{eq:lower_bound}
\end{equation}
where $p_{\mathcal{N}}(x \,;\, m,\sigma)$, $x \in \mathbb{R}$, denotes the density of a Gaussian distribution with mean $m$ and standard deviation
$\sigma$. In terms of an outlier model, this corresponds to a 20\% contamination ratio and nominals
\begin{equation}
  \label{eq:nominals}
  \begin{aligned}
    p_0(x) & = p_{\mathcal{N}}(x \,;\, -1,2), \\
    p_1(x) & = p_{\mathcal{N}}(x \,;\, 1,2).
  \end{aligned}
\end{equation}

The three types of robust tests are obtained by either \emph{clipping}, \emph{censoring} or \emph{compressing} the nominal test statistic.

\begin{figure}[!t]
  \centering
  \subfloat[Clipping]{\includegraphics{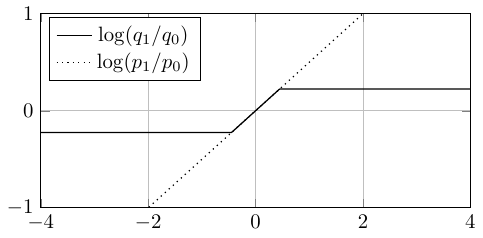} \label{sfig:clipping}} \\[0.25cm]
  \subfloat[Censoring]{\includegraphics{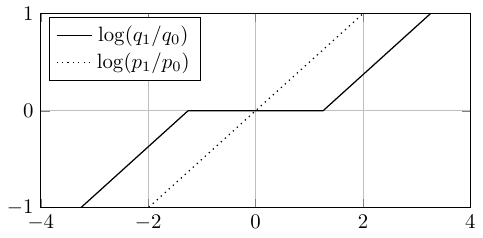} \label{sfig:censoring}} \\[0.25cm]
  \subfloat[Compression]{\includegraphics{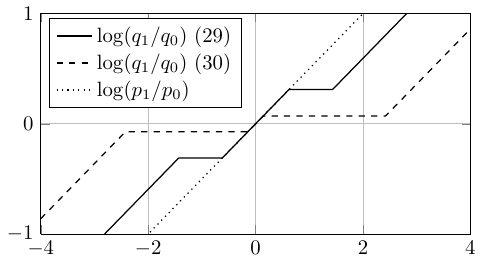} \label{sfig:compression}}
  \caption{Three different types of robust tests resulting from different band models. The least favorable densities are denoted by $q_0,q_1$, the
           nominals, which are given in \eqref{eq:nominals}, by $p_0,p_1$. The lower density bounds for all cases are given in \eqref{eq:lower_bound}.
           The upper bounds for the clipping case (a) are given in \eqref{eq:upper_bound_clipping}, for the censoring case (b) in
           \eqref{eq:upper_bound_censoring}, and for the compression case~(c) in \eqref{eq:uppper_bound_tight} and \eqref{eq:uppper_bound_loose},
           respectively.} 
  \label{fig:band_model_types}
\end{figure}

1) \emph{Clipping:} For very large upper bounds $\pmax_0$, $\pmax_1$ the band model reduces to the $\varepsilon$-contamination model and the minimax optimal test becomes a clipped likelihood ratio test. This means that the influence of a single observation is bounded and seemingly very
significant observations are trusted only to a certain extent. An example is shown in Fig.~\!\ref{sfig:clipping}, where the upper bounds have been
chosen as ten times the nominals:
\begin{equation}
  \pmax_0 = 10 \, p_0 \quad \text{and} \quad \pmax_1 = 10 \, p_1. \label{eq:upper_bound_clipping}
\end{equation}

2) \emph{Censoring:} For moderately large upper bounds the band model often results in likelihood ratios that are \emph{censored} around one. An
example with
\begin{equation}
  \pmax_0 = 1.5 \, p_0 \quad \text{and} \quad \pmax_1 = 1.5 \, p_1 \label{eq:upper_bound_censoring}
\end{equation}
is shown in Fig.~\!\ref{sfig:censoring}. This type of robust test can be seen as a kind of opposite of the clipped test. The influence of highly
significant observations is downweighted, but still unbounded. By contrast, observations whose significance is below a certain threshold are ignored
entirely. This behavior of robust tests has been observed before in the context of \emph{model uncertainties}, meaning that the data is not corrupted
by gross outliers, but that the true distributions differ slightly from the nominal model. Examples for such uncertainty classes are the bounded
KL divergence and bounded Hellinger distance classes, whose least favorable distributions have been shown to result in censored
likelihood ratio tests \cite{Levy2008, Gul2013}.

The intuition behind censored likelihood ratio tests is that under model uncertainties an observation needs to have a certain significance in order to
reliably associate it with a hypothesis. Below this level, there is no clear preference and the observation should simply be ignored.

It is worth mentioning that censoring appears at likelihood ratios close to one, but not necessarily exactly one. In fact, the effectiveness
of censoring depends on choosing the likelihood ratio threshold in accordance with the censoring level, otherwise the test is still minimax optimal
with respect to the weighted sum error probability, but highly biased towards one hypothesis.

3) \emph{Compression:} So far, we have covered the cases of no uncertainty (nominal likelihood ratio test), moderate uncertainty (censored
likelihood ratio test), and high uncertainty (clipped likelihood ratio test). In the transition regions between the nominal and the censored, as
well as between the censored and the clipped test, the band model results in a likelihood ratio that has a plateau at some intermediate level. The
examples in Fig.~\!\ref{sfig:compression} illustrate this phenomenon. They have been obtained by choosing
\begin{equation}
  \pmax_0 = 1.2 \, p_0 \quad \text{and} \quad \pmax_1 = 1.2 \, p_1 \label{eq:uppper_bound_tight}
\end{equation}
and
\begin{equation}
  \pmax_0 = 2.5 \, p_0 \quad \text{and} \quad \pmax_1 = 2.5 \, p_1, \label{eq:uppper_bound_loose}
\end{equation}
respectively. It can be seen that the robust test statistic is the nominal likelihood ratio up to a certain threshold, a constant value on an interval
of medium significance and a scaled version (shifted in the log-domain) of the nominal likelihood ratio on regions of high significance. The test
statistics are neither clipped nor censored, but the influence of most observations is reduced---the more significant they are, the more pronounced
the reduction. We hence refer to this as a \emph{compressed} test statistic.

From the limited number of experiments we performed, it seems that compression is the most frequent outcome of a band model. In general, it offers a
good tradeoff between test performance and robustness with respect to outliers as well as model uncertainties. In order to illustrate this, a more
realistic example is discussed in the next section.

\section{Example}
\label{sec:example}

An idea that arises naturally when dealing with band models is to use confidence intervals of density estimates to construct the bands. Even though
this approach has been suggested repeatedly in the literature \cite{Kassam1981,KassamPoor1985}, we are not aware of any experiments. In this section,
we present the results of a simple experiment that can be seen as a proof of concept for combining confidence intervals and band models.

We consider a problem that arises, for example, in spectrum sensing applications \cite{Yucek2009}, such as cognitive radio \cite{Mitola1999}, where a
secondary user has to reliably detect ongoing transmissions of a primary user, in order to opportunistically occupy or free the channel. Since the
secondary users are often equipped with battery-powered devices and the spectrum has to be sensed frequently, low-complexity energy detectors are a
popular choice for this task \cite{Quan2008,Zhang2009,Zhang2010}.

A commonly used signal model in this context is that of a signal with approximately constant power in complex Gaussian noise
\cite{Quan2008,Zhang2009,Zhang2010}. The hypotheses are accordingly given by
\begin{align*}
  \mathcal{H}_0: & \quad X_k = W_k, \\
  \mathcal{H}_1: & \quad X_k = s_k + W_k
\end{align*}
for $k = 1,\ldots,n$, where $s_k \in \mathbb{C}$ denotes the primary user's complex signal and the $W_k$ denote independently distributed circular
symmetric zero-mean complex Gaussian random variables with variance $\sigma_{\text{W}}^2$. The distribution of $\lvert X_k \rvert^2$ under either
hypothesis is given by \cite{Quan2008}
\begin{equation}
  \begin{aligned}
    \mathcal{H}_0: & \quad \frac{\lvert X_k \rvert^2}{\sigma_{\text{W}}^2} \sim \chi_2^2(0), \\
    \mathcal{H}_1: & \quad \frac{\lvert X_k \rvert^2}{\sigma_{\text{W}}^2} \sim \chi_2^2(\sigma_{\text{S}}^2/\sigma_{\text{W}}^2),
  \end{aligned} \label{eq:hypotheses_example}
\end{equation}
where $\sigma_{\text{S}}^2 = \lvert s_k \rvert^2$ and $\chi_2^2(\zeta)$ denotes the noncentral chi-square distribution with two degrees of freedom and
non-centrality parameter $\zeta$.

\begin{figure}[!t]
  \centering
  \includegraphics{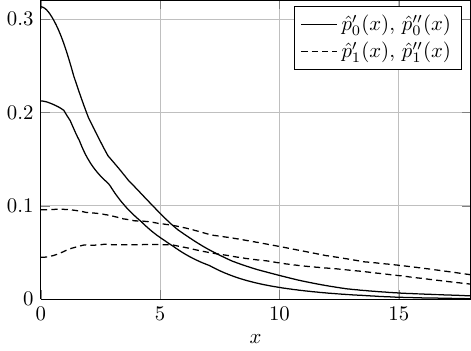}
  \caption{Density bands estimated under $\mathcal{H}_0$ and $\mathcal{H}_1$, respectively, using $400$ samples and $500$ bootstrap instances of a gamma-kernel density estimator.}
  \label{fig:kde_example_bands}
\end{figure}

Often, $\sigma_{\text{S}}^2$ and $\sigma_{\text{W}}^2$ are assumed to be known and constant. In practice, however, these quantities may rapidly vary
with time \cite{Shen2008}. For this example, we assume that $\sigma_{\text{W}}^2 \in [1,2]$ and $\sigma_{\text{S}}^2  \in [4,10]$, which corresponds
to a signal-to-noise ratio $\sigma_{\text{S}}^2 / \sigma_{\text{W}}^2$ between approximately $3\,$dB and $10\,$dB. In this case, the least favorable
distributions are defined by the lowest possible signal-to-noise ratio, i.e., $\sigma_{\text{S}}^2 / \sigma_{\text{W}}^2 = 2$, and can be shown to
be given by
\begin{equation}
  q_0(x) = \frac{1}{\sigma_{\text{W}}^2} p_{\chi^2_2(0)}\left( \frac{x}{\sigma_{\text{W}}^2}\right)
\label{eq:q0_example}
\end{equation}
and 
\begin{equation}
  \label{eq:q1_example}
  q_1(x) = \frac{1}{\sigma_{\text{W}}^2} p_{\chi^2_2(2)}\left( \frac{x}{\sigma_{\text{W}}^2}\right),
\end{equation}
where $p_{\chi^2_2(\zeta)}$ denotes the probability density function corresponding to $\chi^2_2(\zeta)$. Obviously, this model is highly simplified.
The purpose of introducing it here is to investigate how well the least favorable densities \eqref{eq:q0_example} and \eqref{eq:q1_example} can be
reproduced without any knowledge of the model and the signal-to-noise ratio, but only by means of training data and the band model. The procedure
we follow is stated below:
\begin{enumerate}
  \item Take $N_0$ ($N_1$) samples under hypothesis $\mathcal{H}_0$ ($\mathcal{H}_1$).
  \item Calculate density estimates $\hat{p}_0$ and $\hat{p}_1$.
  \item Calculate confidence bands $\hat{p}'_0,\hat{p}''_0$ and $\hat{p}'_1,\hat{p}''_1$.
  \item Calculate $\hat{q}_0,\hat{q}_1$ using the confidence intervals as density bands.
\end{enumerate}
For the experiment, $N_0 = N_1 = 400$ samples were generated under each hypothesis, assuming that noise and signal power vary with every sample and
are uniformly distributed over the respective interval. For the density estimation, we used the second of the two kernel density estimators detailed
in \cite{Chen2000}. It is tailored for densities on the nonnegative reals and uses gamma kernels. The bandwidth was determined via least-squares cross
validation. The estimator was then applied to $500$ data sets that were bootstraped from the original data \cite{Zoubir2007}. The pointwise maximum
and minimum of the corresponding density estimates were used as confidence intervals. The resulting density bands are shown in
Fig.~\ref{fig:kde_example_bands}.

The least favorable densities for this band model were calculated using Algorithm 1 with $\alpha = 0$. The termination criterion suggested in Section
\ref{sec:computation} was used with the norm chosen to be the supremum norm and a tolerance of $10^{-6}$. Convergence was reached after three
iterations. Moreover, a bisection algorithm was used to determine the root of $g_0$ and $g_1$ on the interval $[0,2^{10}]$ so that the bisection
is guaranteed to terminate after at most $10$ iterations.

\begin{figure}[!t]
  \centering
  \includegraphics{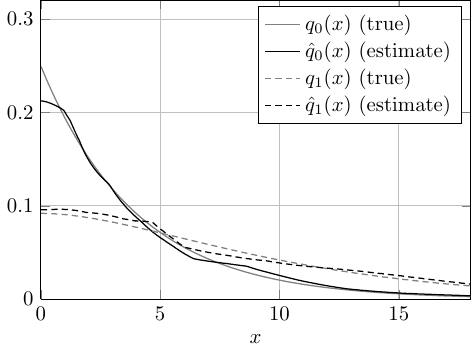}
  \caption{Least favorable densities for the model given in \eqref{eq:hypotheses_example} and least favorable densities for the band model shown in Fig.~\ref{fig:kde_example_bands}.}
  \label{fig:kde_example_lfds}
\end{figure}

Fig.~\ref{fig:kde_example_lfds} depicts the true least favorable densities and the ones estimated by means of the band model.
The gray lines correspond to the former, the black lines to the latter. The deviations of the estimates from the exact solution is clearly visible.
However, given that a rather straightforward estimation approach and no \emph{a priori} knowledge about the underlying distributions was used, the
resemblance is reasonably close under both hypotheses.

\begin{figure}[!t]
  \centering
  \includegraphics{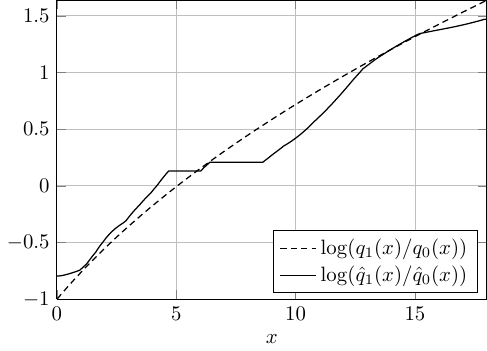}
  \caption{Log-likelihood ratios of the least favorable densities for the model given in \eqref{eq:hypotheses_example} and of the least favorable densities for the band model shown in Fig.~\ref{fig:kde_example_bands}.}
  \label{fig:kde_example_llr}
\end{figure}

The same holds true for the log-likelihood ratio depicted in Fig.~\ref{fig:kde_example_llr}. The estimated robust test statistic is slightly more
conservative, but approximately follows the optimal shape. Two intervals of constant likelihood ratio can be identified between $x=4$ and $x=8$. The
fact that these intervals are close to each other means that the robust testing strategy is closer to censoring than to clipping. This is in line
with the uncertainty model used in this example, which includes varying signal powers, but no gross outliers.

The example demonstrates how the band model can be used in a purely data-driven manner to obtain robust test statistics that adapt to the
underlying distributions and at the same time have a well-defined optimality property. We believe that our results help to better understand the
resulting test and simplify its implementation by offering an efficient and generic algorithm to calculate least favorable densities for any
numerically specified density bands.

\appendices

\section{Proof of Theorem \ref{th:lfds_characterization_3}}
\label{apx:proof_lfds_characterization}

The proof of Theorem \ref{th:lfds_characterization_3} is given in two steps. First, we show that \eqref{eq:lfds_characterization_2} in Theorem
\ref{th:lfds_characterization_2} defines, up to some normalization constant, an $f$-divergence between $P_0$ and $P_1$. Second, it is shown that
a pair of distributions that satisfies Theorem \ref{th:lfds_characterization_3} minimizes all $f$-divergences over $\mathcal{P}_0 \times
\mathcal{P}_1$ and hence satisfies Theorem \ref{th:lfds_characterization_2}.

An $f$-divergence $D_{f}(P_0||P_1)$ between two distributions $P_0$ and $P_1$ is defined via a convex function $f\vcentcolon \mathbb{R}_+ \to
\mathbb{R}$ with $f(1) = 0$ and
\begin{equation}
  \label{eq:f-divergence}
  D_{f}(P_0||P_1) = \smashoperator{\int_{\{p_1 > 0\}}} f\left(\frac{p_0}{p_1}\right) \, \dint P_1 + f'(\infty)P_0[p_1=0],
\end{equation}
where
\begin{equation*}
  f'(\infty) \vcentcolon= \lim_{s \to \infty} \frac{f(s)}{s}.
\end{equation*}
The second term in \eqref{eq:f-divergence} is often omitted under the assumption that $P_1$ dominates $P_0$. In \cite{Osterreicher1993}, it is shown
that every $f$-divergence can equivalently be defined via a convex function $\phi \vcentcolon [0,1] \to \mathbb{R}$ with $\phi(0.5) = 0$ and
\begin{equation}
  D_{\phi}(P_0||P_1) = \int \phi\left(\frac{p_0}{p_0+p_1}\right) d(P_0+P_1). \label{eq:f-div_alternative}
\end{equation}
Moreover, given either $\phi$ or $f$, the complementing function is uniquely determined by
\begin{align}
  f(s)    &= (s+1) \, \phi\left(\frac{1}{s+1}\right), \; s \in \mathbb{R}_+, \quad \text{and} \notag \\
  \phi(s) &= s \, f\left(\frac{1-s}{s}\right), \; s \in [0,1], \label{eq:def_phi}
\end{align}
respectively. Therefore, with $\phi(s) = \Psi(s)-\Psi(0.5)$,
\begin{equation*}
  H_{\Psi}(P_0,P_1) = D_{\phi}(P_0||P_1) + \Psi(0.5)
\end{equation*}
defines an $f$-divergence with constant offset $\Psi(0.5)$, which is independent of $P_0$ and $P_1$.

The last step to prove Theorem \ref{th:lfds_characterization_3} is to note that every $f$-divergence can be written as \cite{Guntuboyina2014}
\begin{equation*}
  D_{f}(P_0||P_1) = c_f - \int L_{\lambda}(P_0,P_1) \, \dint \nu_f(\lambda),
\end{equation*}
where $c_f$ is a constant and $\nu_f$ is a nonnegative measure on $[0,\infty)$, both depending only on $f$. Therefore, if $(Q_0,Q_1)$ maximizes
$L_{\lambda}$ over $\mathcal{P}_0 \times \mathcal{P}_1$ for all $\lambda \geq 0$, it minimizes at the same time all $f$-divergences and in turn
satisfies Theorem \ref{th:lfds_characterization_2}.

\section{Proof of Theorem \ref{th:upper_bound}}
\label{apx:proof_upper_bound}

The proof of Theorem \ref{th:upper_bound} is a straightforward application of Lagrangian duality. The optimization problem at hand is
\begin{equation*}
  \max_{p_0,p_1} \; \int \min \{p_0,\lambda p_1\} \, \dint \mu \quad \text{s.t.} \quad \begin{gathered}[t] 
                                                                                         \int p_i \, \dint \mu = 1, \\
                                                                                         \pmin_i \leq p_i \leq \pmax_i,
                                                                                       \end{gathered}
\end{equation*}
for $i=0,1$. Replacing the minimum under the integral with two additional constraints yields
\begin{equation*}
  \max_{r,p_0,p_1} \; \int r \, \dint \mu \quad \text{s.t.} \quad \begin{gathered}[t]
                                                                    \int p_i \, \dint \mu = 1, \\
                                                                    \pmin_i \leq p_i \leq \pmax_i, \\
                                                                    r \leq p_0, \; r \leq \lambda p_1.
                                                                  \end{gathered}
\end{equation*}
The Lagrangian dual of this problem is
\begin{equation*}
  \min_{s,t,u,v} J(s,t,u,v) \quad \text{s.t.} \quad s,t,u \geq 0,
\end{equation*}
where
\begin{align}
    J(s,t,u,v) ={}& \begin{aligned}[t]
                      \max_{r,p_0,p_1} \bigg\{ \int & r(1-s_0-s_1) \\
                                               &+ \phantom{\lambda} p_0( s_0 + t_0 - u_0 - v_0) \\
                                               &+ \lambda  p_1( s_1 + t_1 - u_1 - v_1) \, \dint \mu \bigg\}
                    \end{aligned} \notag \\
     & + \int (w_0 + \lambda w_1) \, \dint \mu + v_0 + \lambda v_1 \label{eq:def_J}
\end{align}
and we introduced the auxiliary functions
\begin{equation*}
  w_i = u_i \pmax_i - t_i \pmin_i, \quad i=0,1.
\end{equation*}
The real variables $v_i \in \mathbb{R}$ and the real-valued nonnegative functions $s_i,t_i,u_i\colon \Omega \to [0,\infty)$, $i = 0,1$, are
Lagrangian multipliers.\footnote{The multipliers are associated with the constraints as follows:
\begin{equation*}
  \underbrace{r \leq p_0}_{s_0}, \quad \underbrace{r \leq \lambda p_1}_{s_1}, \quad \underbrace{\pmin_i \leq p_i}_{t_i}, \quad
  \underbrace{p_i \leq \pmax_i}_{u_i}, \quad  \underbrace{ \int p_i \, \dint \mu = 1}_{v_i}.
\end{equation*}
Apart from $s_1$, the multipliers associated with $p_1$ have been scaled by $\lambda$ in \eqref{eq:def_J}, i.e., $v_1 \leftarrow
\lambda v_1$, $u_1 \leftarrow \lambda u_1$, and $t_1 \leftarrow \lambda t_1$.}
By strong duality,
\begin{equation}
  \label{eq:dual_bound}
  L_{\lambda}(Q_0,Q_1) \leq J(s,t,u,v) 
\end{equation}
for all $v$ and $s,t,u \geq 0$. Since $J$ is unbounded unless the weighting functions associated with $r$ and $p$ under the integral in
\eqref{eq:def_J} are zero almost everywhere, the dual problem becomes
\begin{gather}
  \min_{s,t,u,v} \; \int w_0 + \lambda w_1 \, \dint \mu + v_0 + \lambda v_1 \label{eq:dual_problem} \\
  \text{s.t.} \quad s,t,u \geq 0, \quad s_0 + s_1 = 1, \quad s+t-u-v = 0. \notag
\end{gather}
Substituting $t = u+v-s \geq 0$ yields
\begin{equation*}
  w_i = u_i(\pmax_i-\pmin_i) + (s_i-v_i)\pmin_i \quad \text{with} \quad u_i \geq s_i-v_i.
\end{equation*}
Since the objective \eqref{eq:dual_problem} is nondecreasing in $u$, the last constraint holds with equality whenever $s-v$ is positive. We can
hence write
\begin{equation*}
  u_i = \max\{s_i-v_i,0\}
\end{equation*}
so that
\begin{equation*}
  \begin{split}
    w_i &= u_i(\pmax_i-\pmin_i) + (s_i-v_i) \pmin_i \\
        &= \max\{s_i-v_i,0\}(\pmax_i-\pmin_i) + (s_i-v_i)\pmin_i  \\
        &= \max\{s_i-v_i,0\} \pmax_i + \min\{s_i-v_i,0\} \pmin_i.
  \end{split}
\end{equation*}
By \eqref{eq:dual_bound}, any feasible combination of dual variables provides an upper bound on $L_{\lambda}$. Assuming that
\begin{equation*}
  v_0,v_1 \in [0,1] \quad \text{and} \quad s_0,s_1 \in \{0,1\}
\end{equation*}
yields
\begin{equation*}
  w_i = \begin{cases}
          (1-v_i)\pmax_i, & s_i = 1 \\
          -v_i \pmin_i,   & s_i = 0
        \end{cases}
\end{equation*}
so that
\begin{align}
  w_0 + \lambda w_1 &\geq \min \{ (1-v_0) \pmax_0 - \lambda v_1 \pmin_1 \,,\, \lambda (1-v_1) \pmax_1 - v_0 \pmin_0 \} \notag \\
                    &= \min \{ \hat{q}_0 \,,\, \lambda \hat{q}_1 \} - v_0 \pmin_0 - \lambda v_1 \pmin_1 \label{eq:integrand_upper_bound}
\end{align}
with $\hat{q}_i$ defined in Theorem \ref{th:upper_bound}. Substituting \eqref{eq:integrand_upper_bound} back into \eqref{eq:dual_problem} yields
the upper bound
\begin{align*}
  L_{\lambda}(Q_0,Q_1) &\leq \begin{aligned}[t]
                                \int & \min \{ \hat{q}_0 \,,\, \lambda \hat{q}_1 \} - v_0 \pmin_0 - \lambda v_1 \pmin_1 \, \dint \mu \\
                                & + \, v_0 + \lambda v_1 
                             \end{aligned} \\
                       &= \int \min \{ \hat{q}_0 \,,\, \lambda \hat{q}_1 \} \, \dint \mu + \varepsilon_0 v_0 + \lambda \varepsilon_1 v_1 .
\end{align*}

\section{Proof of Lemma \ref{lm:root_f}}
\label{apx:proof_root_f}

We only detail the proof for $g_0$, the one for $g_1$ follows analogously. By inspection, $g_0$ is nondecreasing in $c_0$. Moreover, since
$p_0$ and $p_1$ both integrate to one,
\begin{equation*}
  g_0(c_0+\Delta c_0 \,;\, \alpha p_0+p_1) \leq g_i(c_0 \,;\, \alpha p_0 + p_1) + (1+\alpha) \Delta c_0
\end{equation*}
for all $\Delta c_0 \geq 0$ so that $g_0$ is continuous for all finite $\alpha$. Finally,
\begin{equation*}
  g_0(0 \,;\, \alpha p_0+p_1) = \int \pmin_0 \, \dint \mu - 1 \leq 0
\end{equation*}
and
\begin{equation*}
  g_0(\tfrac{1}{\alpha} \,;\, \alpha p_0+p_1) \geq \int p_0 \, \dint \mu - 1 = 0
\end{equation*}
so that
\begin{equation*}
  g_0(c_0 \,; \alpha p_0+p_1) = 0
\end{equation*}
for some $0 \leq c_0^* \leq \tfrac{1}{\alpha}$.

\section{Proof of Convergence of Algorithm 1}
\label{apx:proof_convergence_alg1}

The following corollary will be useful for the proof.
\begin{corollary} \label{cor:f-div_minimizer}
  Let $P \in \Pmu$. If a distribution $Q^*$ with density
  \begin{equation*}
    q^* = \min \{ q'' \,,\, \max \{ c \, p \,,\, q' \} \}
  \end{equation*}
  exists, it is unique and jointly minimizes
  \begin{equation*}
    \D{f}{P}{Q} \quad \text{and} \quad \D{f}{Q}{P}
  \end{equation*}
  among all $Q \in \mathcal{Q}^= = \{ Q \in \Pmu: q'' \leq q \leq q'\}$ and for all convex functions $f$.
\end{corollary}

Corollary~\ref{cor:f-div_minimizer} can be shown as follows: By Theorem~\ref{th:band_lfds}, the pair $(P,Q^*)$ is least favorable in the sense of
Theorem~\ref{th:lfds_characterization_3} if either $\Pcor_0 = \{P\}$ or $\Pcor_1 = \{P\}$ and $\alpha = 0$. From the proof of
Theorem~\ref{th:lfds_characterization_3}, it further follows that $(P,Q^*)$ minimizes all $f$-divergences over the set of feasible distributions
$\mathcal{Q}^=$. Finally, $Q^*$ has to be unique since $\D{f}{P}{Q}$ and $\D{f}{Q}{P}$ can be chosen to be strictly convex in $Q$.

Let us first consider the case when $\alpha = 0$ is a feasible choice. It then follows from Corollary \ref{cor:f-div_minimizer} that for every
$f$-divergence
\begin{equation}
  \begin{split}
    \D{f}{Q_0^n}{Q_1^n} &= \min_{P \in \Pcor_1} \D{f}{Q_0^n}{P} \notag \\
                        &\leq \D{f}{Q_0^n}{Q_1^{n-1}} \notag \\
                        &= \min_{P \in \Pcor_0} \D{f}{P}{Q_1^{n-1}} \notag \\
                        &\leq \D{f}{Q_0^{n-1}}{Q_1^{n-1}} 
  \end{split} \label{eq:monotone_sequence}
\end{equation}
so that $\D{f}{Q_0^n}{Q_1^n}$ is a monotonically nonincreasing sequence of nonnegative real numbers that is guaranteed to converge by the monotone
convergence theorem \cite[Theorem 1.26]{Rudin1987}, i.e,
\begin{equation*}
  \D{f}{Q_0^{n-1}}{Q_1^{n-1}} = \D{f}{Q_0^n}{Q_1^{n-1}} = \D{f}{Q_0^n}{Q_1^n}
\end{equation*}
for $n \to \infty$. Since $Q_0^n$ and $Q_1^n$ are unique minimizers of $\D{f}{Q_0^n}{Q_1^{n-1}}$ and $\D{f}{Q_0^n}{Q_1^n}$, respectively, this
implies that in the limit $Q_0^{n-1} = Q_0^n$ and $Q_1^{n-1} = Q_1^n$.

For $\alpha > 0$, the proof is similar. Instead of general $f$-divergences, we consider the total variation distance, which is obtained by
choosing $f(s) = f_{\text{TV}}(s) = \lvert 1-s \rvert$ and can be written as 
\begin{equation*}
  \D{f_{\text{TV}}}{P}{Q} = \int \lvert p-q \rvert \, \dint \mu = \lVert p-q \rVert,
\end{equation*}
where $\lVert \, \cdot \, \rVert$ denotes the $L^1$-norm. Note that the total variation distance is symmetric and satisfies the triangle inequality.
With $\pi := \tfrac{\alpha}{1+\alpha} \in (0,1)$, it is the case that
\begin{equation*}
  \begin{split}
  \lVert q_0^n-q_1^n \rVert &\stackrel{\mathclap{(\RN{1})}}{\leq} \begin{multlined}[t][0.7\columnwidth]
                                                                    \lVert q_0^n-\pi q_0^n - (1-\pi) q_1^{n-1} \rVert \\
                                                                    + \lVert q_1^n-\pi q_0^n - (1-\pi) q_1^{n-1} \rVert
                                                                  \end{multlined} \\
    &\stackrel{\mathclap{(\RN{2})}}{\leq} \begin{multlined}[t][0.7\columnwidth]
                                            (1-\pi)\lVert q_0^n - q_1^{n-1} \rVert \\
                                            + \lVert q_1^{n-1}-\pi q_0^n - (1-\pi) q_1^{n-1} \rVert
                                          \end{multlined} \\
    &= (1-\pi)\lVert q_0^n - q_1^{n-1} \rVert + \pi \lVert q_0^n - q_1^{n-1} \rVert \\
    &= \lVert q_0^n - q_1^{n-1} \rVert \\
    &\stackrel{\mathclap{(\RN{3})}}{\leq} \begin{multlined}[t][0.7\columnwidth]
                                            \lVert q_0^n - \pi q_0^{n-1} - (1-\pi) q_1^{n-1} \rVert \\
                                            + \pi \lVert q_0^{n-1} - q_1^{n-1} \rVert
                                          \end{multlined} \\
    &\stackrel{\mathclap{(\RN{4})}}{\leq} \begin{multlined}[t][0.7\columnwidth]
                                            \lVert q_0^{n-1} - \pi q_0^{n-1} - (1-\pi) q_1^{n-1} \rVert \\
                                            + \pi \lVert q_0^{n-1} - q_1^{n-1} \rVert
                                          \end{multlined} \\
    &= \lVert q_0^{n-1} - q_1^{n-1} \rVert,
  \end{split}
\end{equation*}
where $(\RN{2})$ and $(\RN{4})$ follow from Corollary \ref{cor:f-div_minimizer} by construction of $q_1^n$ and $q_0^n$, respectively. Hence,
\begin{equation*}
  \lVert q_0^{n-1} - q_1^{n-1} \rVert = \lVert q_0^n - q_1^{n-1} \rVert = \lVert q_0^n - q_1^n \rVert,
\end{equation*}
for $n \to \infty$. Finally, the triangle inequalities $(\RN{1})$ and $(\RN{3})$ only hold with equality if $Q_1^{n-1} = Q_1^n$ and $Q_0^{n-1} =
Q_0^n$, which concludes the proof.

%
%

\ifCLASSOPTIONcaptionsoff
  \newpage
\fi



\bibliographystyle{IEEEtran}
\bibliography{bibliography}
%
%
%

%

%
%
%





\end{document}